\documentclass[aps,prb,twocolumn,showpacs,superscriptaddress,groupedaddress]{revtex4-2}
\usepackage{amsmath}
\usepackage{amssymb}
\usepackage{graphicx}
\usepackage{bm}
\usepackage{float}
\usepackage{xcolor}
\usepackage{times}
\usepackage{notes2bib}
\usepackage{verbatim}
\usepackage[none]{hyphenat}
\usepackage{braket}
\usepackage{dcolumn}

\newcommand{\w}{\omega}

\begin{document}
\title{Insulating moir\'e homobilayers lack a threefold symmetric second harmonic generation}
\author{Luis Enrique Rosas-Hernandez,$^1$ Jose Luis Cabellos,$^2$ Angiolo Huam\'an,$^1$ Bernardo Mendoza,$^{3,4*}$ and Salvador Barraza-Lopez$^{1}$}
\email{bms@cio.mx, sbarraza@uark.edu}
\affiliation{1. Department of Physics. University of Arkansas. Fayetteville, AR 72701, USA.\\
2. Universidad Polit\'ecnica de Tapachula. Tapachula, Chiapas, C.P.~30830, M\'exico.\\
3. Centro de Investigaciones en \'Optica, A.C.  Le\'on, Guanajuato, C.P.~37150, M\'exico.\\
4. Max Planck Institute for the Structure and Dynamics of Matter, Luruper Chaussee 149, Hamburg, Germany.}

\begin{abstract}
{Atoms within moir\'e bilayers relax in-plane to minimize elastic energy [{\em e.g.}, Cazeaux {\em et al.}, {\em J. Elast.} {\bf 154}, 443 (2023)]; such relaxation brings their space group symmetries down to P1.} Here, the {\em ab initio} second harmonic generation (SHG) of twisted { and atomistically optimized} hBN bilayers was determined at four twist angles ($\theta=38.21^{\circ}$, $60.00^{\circ}$, $73.17^{\circ}$, and $98.21^{\circ}$) and for three displacements $\boldsymbol{\tau}$ measured away from the ground state $AA^{\prime}$ configuration. { All moir\'e bilayers have a P1 space symmetry after structural optimization. This situation is quite different to monolayers with hexagonal lattices, which retain a three-fold symmetry. We point out that the actual symmetries of the SHG reported for hBN bilayers on two experimental works do not coincide with the sixfold symmetric theoretical profiles they provide [either $\sin^2(3\phi)$ or $\cos^2(3\phi)$], and show that the} {\em intrinsic} low structural symmetry of { (atomically optimized)} hBN bilayer moir\'es can in fact be read out from experimental SHG intensity profiles--which are tunable by $\theta$ and by the frequency of light $\omega$: { The SHG is {\em most definitely not} sixfold-symmetric because moir\'es do not retain a three-fold symmetry.} Furthermore, an {\em extrinsic} twofold symmetry of the SHG emission is realized by tilting the pump by an angle $\alpha$ away from the 2D material's normal, regardless of $\theta$ and $\omega$. The design of in-plane and ultrathin sources of SHG with low symmetry could be useful for the eventual creation of entanglement sources from 2D materials.
\end{abstract}
\maketitle

\section{Introduction}
Highlighting the promise of 2D materials for nonlinear and quantum optics, transition metal dichalcogenide and hBN \textit{monolayers} have peak SHG intensities at least five times larger than those reported in bulk non-centrosymmetric insulators and semiconductors \cite{Malard2013,Kumar2013, Qian2017,Li2013}. {  Numerical and theoretical advances on the SHG response of those monolayers include the development of techniques that include excitonic effects either through a real-time approach--and whose main contribution is to double the first SHG peak intensity \cite{Gruning2014}, through a two-band model reliant on the orthogonality between $\pi-$ and $\sigma-$states on monolayers with a hexagonal lattice \cite{Pedersen2015}, or using tight-binding approaches \cite{PhysRevB.89.235410}. Multiple studies on the non-linear responses of--threefold symmetric--graphene are summarized in Ref.~\cite{GLAZOV2014101} as well.}

Most layered materials retain centers of inversion when thinned down to a bilayer and thus lack a SHG response, but a SHG intensity even larger than that found in hBN monolayers was measured in few-layer hBN as early as 2013 \cite{Kim2013}. The authors argued that its origin was the breaking of inversion symmetry by a $\theta=60^{\circ}$ twist among monolayers--for a { {\em still three-fold symmetric}} $AB$ or $BA$ ($AC$) non-centrosymmetric stacking--during growth \cite{Kim2013}. { The excitonic response of a moir\'e bilayer without additional atomistic optimization was contributed in Ref.~\cite{Taboada2023} which, as it will be soon discussed, still has an artificial three-fold symmetry.}

\begin{figure}[tb]
\includegraphics[width=\columnwidth]{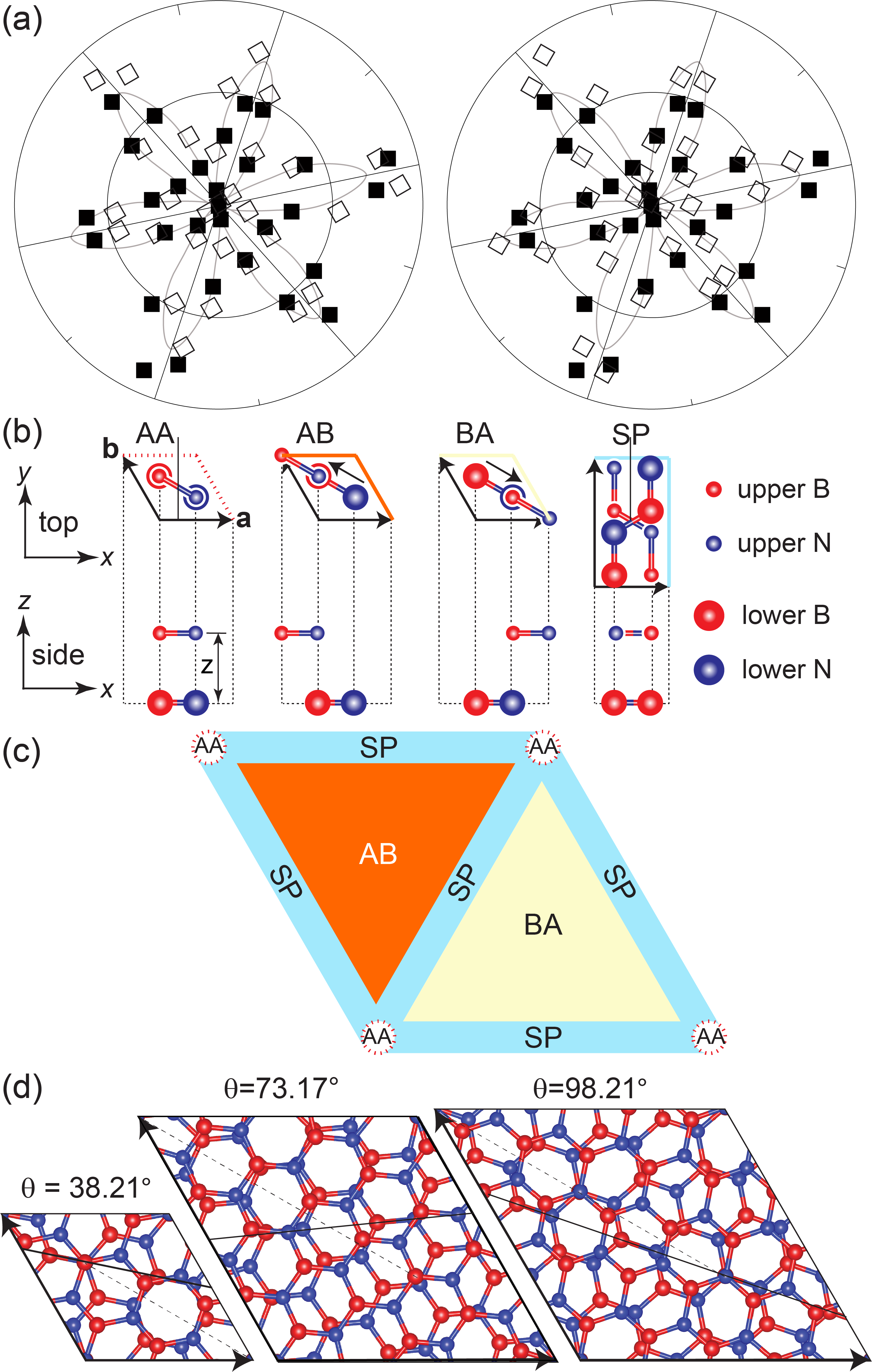}
\caption{(a) The experimental SHG intensity from Ref.~\cite{Kim2013} (black) was rotated by $120^{\circ}$ (left subplot) and $240^{\circ}$ (right subplot) to emphasize a missing threefold symmetry (open squares). It lacks a sixfold symmetry, too. (b) $AA$, $AB$, $BA$, and $SP$ bilayers ($\theta=60^{\circ}$ and three $\boldsymbol{\tau}$'s were employed). (c) Schematic placement of the atomistic configurations in (b) on a large-area moir\'e. (d) Twisted hBN bilayers for $\theta=38.21^{\circ}$, 73.17$^{\circ}$, and 98.21$^{\circ}$. Solid lines in (b) and (d) indicate the armchair direction of the bottom monolayer. { The dashed lines along the main diagonals in subplot (d) help visualize how much atoms at locations 1/3 or 2/3 along said diagonals moved away to reduce in-plane strain.}\label{fig:Fig1}}
\end{figure}

{ {\em This manuscript is about how structural symmetry affects a SHG response}: Regardless of the level of the theory, structural symmetries determine nonzero matrix elements, and by extension the symmetries of the SHG response. Therefore, all works discussing monolayers with a honeycomb lattice in previous two paragraphs will yield a six-fold symmetric [$\sin^2(3\phi)$ or the equivalent $\cos^2(3\phi)$, with $\phi$ the linear polarization angle] SHG response. On the other hand, given their atomistic reconstruction, it is incorrect to assign a sixfold symmetry to the SHG response of moir\'e bilayers, as done in Refs.~\cite{Kim2013} and \cite{Yao2021}.} As reproduced in Fig.~\ref{fig:Fig1}(a), the ascribed sixfold-symmetric angular dependence [$\sin^2(3\phi)$] is inconsistent with the experimental data reported in Ref.~\cite{Kim2013}. { (Given that no error bars were provided, we assume they are of the order of the square features shown in the figure.)}

The experimental difficulty in creating twist angles that are exactly $\theta=60^{\circ}$, along with an ever-present local strain, unavoidably yield moir\'es on twisted bilayers \cite{Vizner2021,Yasuda2021,Woods2021,Moore2021,Engelke}. Along those lines, Yao {\it et al.}~cut a thick hBN crystal in two parts and measured the SHG intensity for $\theta=40.9^{\circ}$, 62.4$^{\circ}$, 79.6$^{\circ}$, and 99.0$^{\circ}$, among other twist angles \cite{Yao2021}. Similar to Ref.~\cite{Kim2013}, they state that the point group of AB hBN bilayers is the (threefold symmetric) 32, and that the SHG intensity is proportional to a {\em sixfold symmetric} $\cos^2(3\phi)$. { If the symmetry assigned to the SHG responses for multiple moir\'e configurations in Ref.~\cite{Yao2021} were true, the purported threefold symmetry would be consistent with the atomistic structures reported in Refs.~\cite{Taboada2023} and \cite{Latil2023}, none of which are atomically relaxed.}

{ Reference  \cite{Latil2023} makes the point that all threefold symmetric moir\'es in hBN bilayers belong to space groups p321, p312, or p3, depending on lattice matching, resulting in only five unit cells. To start with, the lowercase (`p') classification only applies to 2D (paperwall) crystals. The fact that there are two layers calls for a ``bulk'' (`P') space group notation \cite{itc}. Besides this, the authors of that work did not carry out any atomistic optimization, leading to the symmetries they report. In reality though, the in-plane atomistic reconstruction of moir\'es \cite{reconstruction} is well-known by now, and can be understood as follows: a moir\'e is a realization of multiple unit cell configurations, with atoms in the upper monolayer unit cell relatively displaced with respect to the two atoms in the lower unit cell; see Ref.~\cite{Engelke}. Atoms within each monolayer move in-plane to avoid the high energy local bilayer configurations computed in Ref.~\cite{Zhou2015}; this experimentally-verifiable process has been even written in terms of a functional optimization in Ref.~\cite{JElasticity}.}

{ And so, the core of our argument is that the SHG plots in Refs.~\cite{Kim2013} and \cite{Yao2021} betray a lower symmetry than the one both experimental papers report, and that the observed features can be explained by recourse to the reduced symmetry of moir\'es, which is lower than the one reported in Refs.~\cite{Taboada2023} and \cite{Latil2023}.} The experimental results in Refs.~\cite{Kim2013} and \cite{Yao2021} (both missing {\em ab initio} SHG calculations) beg the question of whether the observed reduced symmetry of the SHG intensity is {\em intrinsic}, or arising from within an experimental setup, {  and we show here that its origin is intrinsic indeed}. Moving beyond a curiosity, a SHG intensity with a properly tuned, axial symmetry could benefit sources of entangled photon pairs by making individual photons travel {\em along opposite directions by design}.

{ This way, this manuscript does not contain new methodologies for the calculation of excitonic effects on moir\'e bilayers (it is based on a single-particle approach for the SHG). It is concerned with how the P1 symmetry of moir\'e homobilayers lowers the symmetry of the SHG, and it also contains the first calculations for SHG in relaxed hBN bilayer moir\'es. We claim that the experimentally reported results confirm the low-symmetry of moir\'es already. To the extent that low-symmetries will remain even when many-body effects are included, we are confident that the lower symmetry of the SHG will stand scrutiny.}

To quantify the claims above, this paper contains {\em ab initio} calculations of the parallel second order optical susceptibility $\chi_{\|}^{(2)}(\theta,\phi,\alpha;-2\omega,\omega,\omega)$ of unstrained { (i.e., atomically relaxed)} hBN bilayers. The calculations permit distinguish intrinsic from extrinsic sources of lowered SHG intensity (the pump's frequency is $\omega$). Twist angles  $\theta=60.00^{\circ}$ (for three values of the in-plane displacement $\boldsymbol{\tau}$), $38.21^{\circ}$, 73.17$^{\circ}$, and 98.21$^{\circ}$, and multiple probe angles $\alpha$ with respect to the bilayer's normal were considered. As a result, the low symmetry observed in experimental data is explained, and a new process to create an axial SHG off 2D materials discovered.

\section{Methods}
Structural optimizations of twisted hBN bilayers were performed using plane-wave density functional theory (DFT) as implemented in the ABINIT package \cite{Gonze2002}. The general gradient approximation (GGA) with Perdew-Burke-Ernzerhof (PBE) exchange-correlation functionals \cite{Perdew1996} and a semiempirical vdw-DFT-D2 correction to account for dispersion forces~\cite{Grimme2006}, and optimized norm-conserving Vanderbilt pseudopotentials \cite{Hamann2017}, were employed. The plane wave cutoff energy was set to 25 Hartree, and Monkhorst-Pack~\cite{Monkhorst1976} $k-$point samplings of  $21\times 21 \times 1$ and $4\times4\times1$ centered at the $\Gamma$ point were used for structures with 4 atoms ($\theta=60.00^{\circ}$) and for moir\'es containing 28 ($\theta=38.21^{\circ}$), 76 (73.17$^{\circ}$) and 84 (98.21$^{\circ}$) atoms, respectively. { Structural optimizations of the moir\'es were carried out with a force tolerance of $10^{-5}$ Ha/Bohr.} The $\chi_{\|}^{(2)}(\theta,\phi,\alpha;-2\omega,\omega,\omega)$ tensor was calculated from the real-space Bloch wavefunctions obtained from DFT calculations with the in-house TINIBA code (see Supplementary Material~\cite{SM} for details). Excitonic effects were not considered, as justified in the preceding section.

\section{Results}
\subsection{Symmetry of hBN bilayers}

Following the naming convention from Ref.~\cite{Mele2010}, ground state hBN bilayers have 4 atoms in their unit cells (u.c.s). The in-plane location of the boron (nitrogen) atom in the lower monolayer is given by $\mathbf{a}/3+2\mathbf{b}/3$ ($2\mathbf{a}/3+\mathbf{b}/3$), with lattice vectors defined as $\mathbf{a} = a(1,0)$ and $\mathbf{b} = a (-1/2,\sqrt{3}/2)$, and a lattice constant $a=2.508$ \AA. The boron (B) and nitrogen (N) atoms in the upper monolayer are obtained by a vertically displaced (by $z$) copy of the lower monolayer (leading to a non-centrosymmetric $AA$ bilayer), and a subsequent swap of chemical elements on the upper monolayer (for a centrosymmetric $AA^{\prime}$ bilayer) [Fig.~\ref{fig:Fig1}(b)]. The $AA$ bilayer can also be obtained from the $AA^{\prime}$ one by a rotation $\theta=60^{\circ}$ with respect to a perpendicular axis passing through a hexagon's center.

Two energetically degenerate configurations labeled $AB$ and $BA$ are obtained by additional in-plane displacements of the upper monolayer by $\boldsymbol{\tau}=(-\bf{a}+\bf{b})/3$ or $\boldsymbol{\tau}=-(-\mathbf{a}+\mathbf{b})/3$, respectively, away from the $AA$ configuration [Fig.~\ref{fig:Fig1}(b)]. A smaller displacement away from the $AA$ configuration [$\boldsymbol{\tau}=(-\mathbf{a}+\mathbf{b})/2$] leads to the { (``saddle point'')} $SP$ bilayer in Fig.~\ref{fig:Fig1}(b); { a notation introduced by Zou and coworkers in 2015 that reflects the location of this structure within the energy landscape of the bilayer upon sliding \cite{Zhou2015}.}

The left column in Table \ref{tab:t1} contains the space and point groups of the $AA$, $AB$ ($BA$), and $SP$ hBN bilayers, and their DFT electronic band gaps~\cite{SM}. The $AA$ bilayer is the most symmetric, while the $SP$ one has a twofold symmetry. Those atomistic arrangements are observed on large-area twisted (moir\'e) hBN bilayers [Fig.~\ref{fig:Fig1}(c)]; the area with an $AA$ configuration is small, and domains $AB$ and $BA$ are separated by $SP$ domain walls \cite{Engelke}.

Reference~\cite{Yao2021} erroneously states that $AB$ hBN bilayers belong to the 32 point group. Indeed, the space group of $AB$ and $BA$ hBN bilayers is P3m1 (space group No. 156) \cite{Hahn2002}, which has six symmetry operations:
\begin{equation*}\label{eq:P3m1}
\begin{aligned}
(1):&\hspace{0.2cm}x,y,z               \hspace{2mm} &&(2):\hspace{0.2cm}\bar{y},x-y,z  \hspace{2mm} &&&(3):\hspace{0.2cm}\bar{x}+y,\bar{x},z \\
(4):&\hspace{0.2cm}\bar{y},\bar{x},z   \hspace{2mm} &&(5):\hspace{0.2cm}\bar{x}+y,y,z  \hspace{2mm} &&&(6):\hspace{0.2cm}x,x-y,z
\end{aligned}
\end{equation*}
where the bar over a coordinate implies a negative sign ({\em e.g.}, $\bar{x}=-x$) and the fractional coordinates are multiplied by lattice vectors to yield Cartesian positions.

No symmetry operation of space group P3m1 places atoms on the opposite monolayer, which permits focusing on $x$ and $y$ only. Concluding the proof, atomic positions remain unchanged for $x=0$ and $y=0$, $x=1/3$ and $y=2/3$, or $x=2/3$ and $y=1/3$, which include all atomic positions depicted in Fig.~\ref{fig:Fig1}(b) for the $AB$ and $BA$ configurations (positions are equivalent with additions by $\pm 1$). The P3m1 space group has $3m$ as its point group \cite{Hahn2002}, {\em not} 32 \cite{Yao2021}.

For the SHG intensity from twisted (moir\'e) hBN bilayers to be sixfold symmetric \cite{Kim2013,Yao2021} at normal incidence ($\alpha=0^{\circ}$), the different areas within a moir\'e [Fig.~\ref{fig:Fig1}(c)] must contribute coherent three-fold symmetric second harmonics. (The presence of $SP$ domain walls with three orientations at $0^{\circ}$, $120^{\circ}$, and $240^{\circ}$ may restore the three-fold symmetry on the SHG response arising from those locations within a moir\'e.) Nevertheless, {\em ab initio} calculations will show that moir\'es have polar SHG intensities with lower symmetry.

\begin{table}[tb]
\centering
\caption{Angle $\theta$, space and point groups, and DFT electronic band gaps for the six hBN bilayers studied. The letter D (I) stands for a direct (indirect) band gap. The bilayers lack a center of inversion.}
\label{tab:t1}
\begin{tabular}{cccc|cccc}
\hline
\hline
$\theta$       & Space      & Point     & Band gap  &$\theta$     & Space & Point     & Band gap       \\
$(^{\circ})$   & group      & group     & (eV)      &$(^{\circ})$ & group & group     & (eV)           \\
\hline
 60 ($AA$)     &P$\bar{6}$m2&$\bar{6}$m2& 3.98 (D)  & 38.21       & P1    & 1         & 4.33 (I)\\
 60 ($AB$,$BA$)& P3m1       &  3m       & 4.46 (I)  & 73.17       & P1    & 1         & 4.21 (I)\\
 60 ($SP$)     & Aem2       & mm2       & 4.22 (I)  & 98.21       & P1    & 1         & 4.35 (I)\\
\hline
\hline
\end{tabular}
\end{table}

Starting from an $AA^{\prime}$ configuration, a commensurate twisted hBN bilayer is constructed by a rotation of the lower and upper monolayers by $-\theta/2$ and $\theta/2$, respectively. Commensuration requires the rotation angle $\theta$ to satisfy \cite{Mele2010}:
\begin{equation}\label{eq:thetam}
\cos{\theta} = -\frac{m^2+n^2-4mn}{2(m^2+n^2-mn)} \text{ for $m$ and $n$ integers.}
\end{equation}

Figure~\ref{fig:Fig1}(d) shows moir\'es after a rotation--by $\theta=38.21^{\circ}$ [$(m,n)=(3,2)$], 73.17$^{\circ}$ [$(m,n)=(5,2)$], and 98.21$^{\circ}$ [$(m,n)=(5,1)$]--and an atomistic optimization. Their lattice constants are $\sqrt{7}a$, $\sqrt{19}a$, and $\sqrt{21}a$, respectively.

{ The authors of Ref.~\cite{Latil2023} claim that there are only five moir\'e configurations available with threefold rotational symmetry, depending of which atoms are at the origin, and at a position 2/3 along the main diagonal. In fact, due to optimization, there is only one space group: $P1$ (complete absence of symmetry), so only before atomistic optimization, our structures can be classified with the BNNB notation \cite{Latil2023}. Table \ref{tab:move} says that the atoms at 2/3 (or 1/3) along the diagonal move slightly away {\em and break threefold symmetry}.}

\begin{table}[tb]
\centering
\caption{ Location of atoms along the main diagonal in terms of supercell lattice vectors $\mathbf{a}$ and $\mathbf{b}$. Note that the atoms along the diagonal are {\em not} at 1/3 or 2/3 along the main diagonal, and hence {\em the moir\'es lack three-fold symmetry}. The third coordinate is to be multiplied by $c=20$ \AA, to get the actual height in calculations.}
\label{tab:move}
\begin{tabular}{cc}
\hline
\hline
$\theta=38.21^{\circ}$\\
\hline
B (0.00000, 0.00000, 0.40000) & N (0.00000, 0.00000, 0.61333) \\
N (0.33675, 0.66687, 0.40000) & B (0.33013, 0.66687, 0.61333) \\
\hline
$\theta=73.17^{\circ}$\\
\hline
B (0.00000, 0.00000, 0.40000) & N (0.00000, 0.00000, 0.61333) \\
N (0.66641, 0.31722, 0.40000) & B (0.66641, 0.34919, 0.61333) \\
\hline
$\theta=98.21^{\circ}$\\
\hline
B (0.00000, 0.00000, 0.40000) & N (0.00000, 0.00000, 0.61333) \\
N (0.66667, 0.33441, 0.40000) & B (0.66667, 0.33226, 0.61333) \\
\hline
\hline
\end{tabular}
\end{table}

This previous observation happens because twisted hBN bilayers reconstruct to reduce (energetically costly) areas with an $AA$ configuration \cite{Engelke,JElasticity} and, as a result, they belong to { space point P1 and to }point group $1$ (right column of Table \ref{tab:t1}, which includes DFT electronic band gaps as well \cite{SM}).

\subsection{Optical second order susceptibility}
\begin{figure}[tb]
\includegraphics[width=\columnwidth]{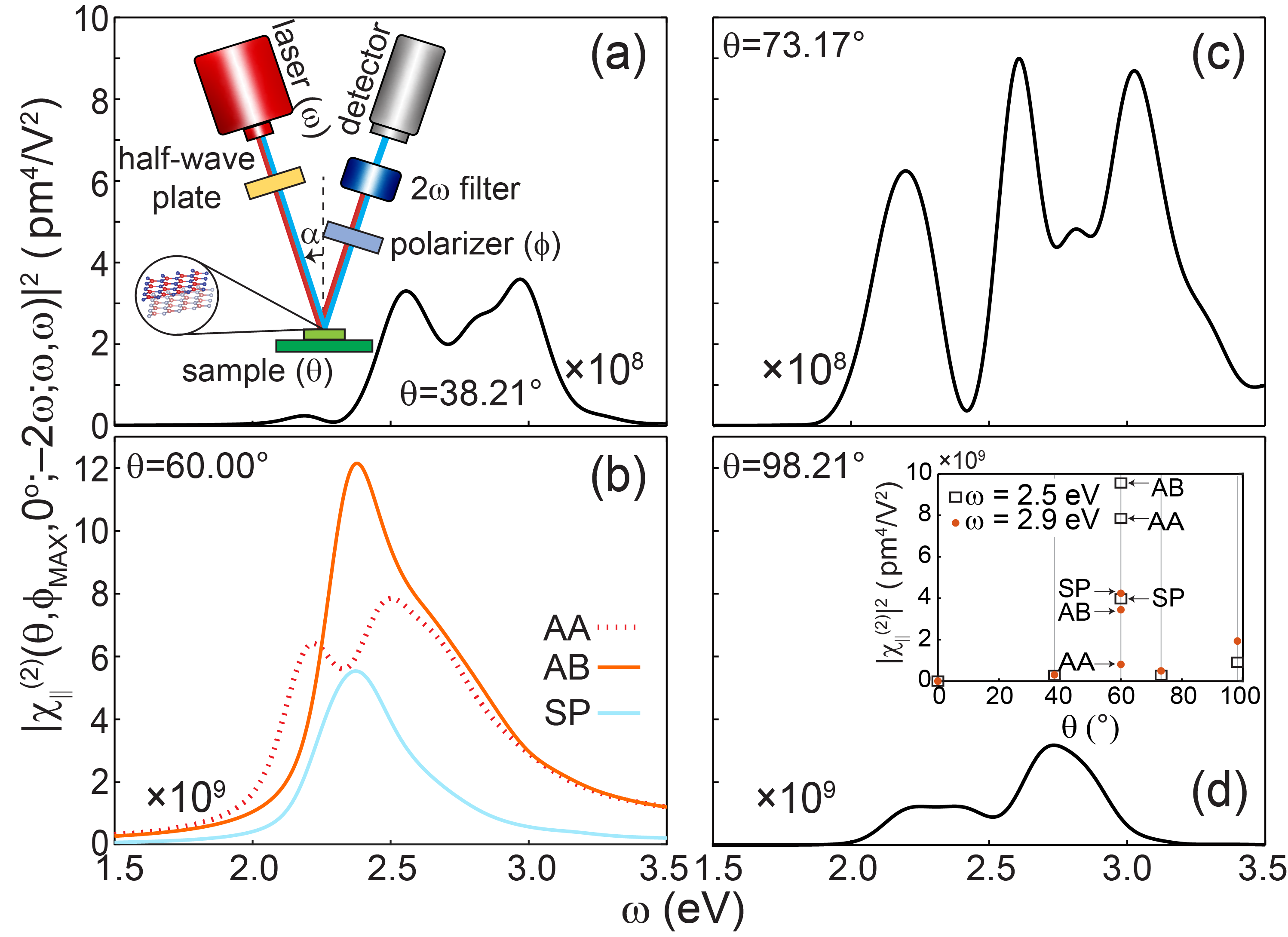}
\caption{
Maximum intensity $|\chi_{\|}^{(2)}(\theta,\phi_{MAX},\alpha=0^{\circ};-2\omega,\omega,\omega)|^2$ {\em versus} $\theta$ and $\omega$. Subplot (a) includes schematics of a reflection setup and the dependencies of $\theta$ on the sampled bilayer, $\omega$ on the probe, and of $\phi$ and $\alpha$ on the polarizer and the relative orientation among sample and probe, respectively. The inset in subplot (d) illustrates the increase in SHG intensity in going from $\theta=0^{\circ}$ to $60^{\circ}$.\label{fig:Fig2}}
\end{figure}

SHG experiments in reflection mode may be performed with a tilted probe by an angle $\alpha$ that influences the intensity profile. Therefore, the SHG intensity is calculated as a function of $\phi$ and $\alpha$ in what follows. Quadrupole contributions to the SHG \cite{Yao2021} are negligible in bilayers, and the second order optical susceptibility tensor $\chi_{ijk}^{(2)}(\theta;-2\w;\w,\w)$ (with $i$, $j$, and $k$ Cartesian components) depends on the stacking through $\theta$ (the dependence on $\boldsymbol{\tau}$ on the 4-atom u.c.s is not explicitly written). As illustrated by an inset in Fig.~\ref{fig:Fig2}(a), it is defined by the relation between the induced second order electric polarization $\mathbf{P}(\theta;2\w)$ and the applied electric field $\mathbf{E}(\phi,\alpha;\w)$~\cite{Boyd}:
\begin{equation*}
	P_i(\theta,\phi,\alpha;2\w) = \chi_{ijk}^{(2)}(\theta;-2\w;\w,\w) E_j(\phi,\alpha;\w) E_k(\phi,\alpha;\w),
\end{equation*}
with $\chi_{ijk}^{(2)}=\chi_{ikj}^{(2)}$, and an implied sum over repeated indices. A tilted linearly-polarized electric field is written as:
\begin{equation*}
\mathbf{E}(\phi,\alpha;\w) = E_0(\omega) (\cos{\phi} \cos{\alpha}, \sin{\phi}, -\cos{\phi}\sin{\alpha}).
\end{equation*}

Writing the  quadratic dependence of $\mathbf{E}$ on $\phi$ and $\alpha$ in the expression for $P_i(\theta,\phi,\alpha;2\w)$ explicitly, one gets:
\begin{eqnarray}\label{eq:chipar}
& \chi_{\|}^{(2)}(\theta,\phi,\alpha;-2\omega,\omega,\omega) \equiv \mathbf{E}\cdot \frac{\mathbf{P}}{E_0^3} =\\
& \chi^{(2)}_{xxx} \cos^{3}{\phi} \cos^{3}{\alpha} + \chi^{(2)}_{yyy} \sin^{3}{\phi} - \chi^{(2)}_{zzz} \cos^{3}{\phi}\sin^{3}{\alpha} \nonumber\\
&+(2\chi^{(2)}_{xxy}+ \chi^{(2)}_{yxx}) \sin{\phi} \cos^{2}{\phi}\cos^{2}{\alpha}\nonumber\\
&+(\chi^{(2)}_{xyy} +2\chi^{(2)}_{yxy}) \sin^{2}{\phi} \cos{\phi} \cos{\alpha} \nonumber \\
&-(2\chi^{(2)}_{xxz}+ \chi^{(2)}_{zxx})\cos^{3}{\phi} \cos^{2}{\alpha} \sin{\alpha}\nonumber \\
&+(\chi^{(2)}_{xzz} +2\chi^{(2)}_{zxz}) \cos^{3}{\phi}  \sin^{2}{\alpha} \cos{\alpha}\nonumber\\
&-(2\chi^{(2)}_{yyz}+ \chi^{(2)}_{zyy}) \sin^{2}{\phi}  \cos{\phi} \sin{\alpha} \nonumber \\
&+(\chi^{(2)}_{yzz} +2\chi^{(2)}_{zyz}) \sin{\phi}   \cos^{2}{\phi} \sin^{2}{\alpha}\nonumber\\
&-(2\chi^{(2)}_{xyz}+ \chi^{(2)}_{yxz}) \sin{\phi}  \cos^{2}{\phi} \sin{\alpha}\cos{\alpha}\nonumber\\
&-(\chi^{(2)}_{yxz} +2\chi^{(2)}_{zxy}) \sin{\phi}  \cos^{2}{\phi} \sin{\alpha}\cos{\alpha},\nonumber
\end{eqnarray}
where the dependence of $\chi^{(2)}_{ijk}$ on $\theta$ and $\omega$ was omitted for brevity. According to Table \ref{tab:t1}, the point group of the $AA$ configuration is $\bar{6}$m2, and it has a single independent nonzero entry on $\chi^{(2)}_{ijk}$ \cite{Boyd}:
\begin{equation}\label{eq:aa}
\chi^{(2)}_{||_{AA}}(\phi,\alpha) = \chi^{(2)}_{yyy} (\sin^{3}{\phi} - 3\sin{\phi}\cos^{2}{\phi}\cos^{2}{\alpha}),
\end{equation}
where the dependence on $\w$ was omitted. Since $AB$ (and $BA$) stacked hBN bilayers belong to the point group 3m, their $\chi^{(2)}_{ijk}$~\cite{Boyd} has four independent, nonzero entries:
\begin{eqnarray}\label{eq:ab}
\chi^{(2)}_{||_{AB}}(\phi,\alpha)= \chi^{(2)}_{yyy} (\sin^{3}{\phi} - 3\sin{\phi}\cos^{2}{\phi}\cos^{2}{\alpha}) \\
 - (2\chi^{(2)}_{xzx} + \chi^{(2)}_{zxx}  ) \sin{\alpha}  \cos{\phi} (\cos^{2}{\phi} \cos^{2}{\alpha} + \sin^{2}{\phi}) \nonumber \\
 -\chi^{(2)}_{zzz} \cos^{3}{\phi} \sin^{3}{\alpha}.\nonumber
\end{eqnarray}
Still relying on Table \ref{tab:t1}, the point group for the $SP$ configuration is mm2, which has five independent non-zero entries for $\chi^{(2)}_{ijk}$~\cite{Boyd}:
\begin{eqnarray}\label{eq:sp}
\chi^{(2)}_{||_{SP}}(\phi,\alpha) = \chi^{(2)}_{yyy} \sin^{3}{\phi} \\
+(2\chi^{(2)}_{xxy} + \chi^{(2)}_{yxx}  ) \sin{\phi}  \cos^{2}{\phi} \cos^{2}{\alpha}\nonumber \\
+(\chi^{(2)}_{yzz} + 2\chi^{(2)}_{zyz}) \sin{\phi}\cos^{2}{\phi} \sin^{2}{\alpha}.\nonumber
\end{eqnarray}

Lacking crystalline symmetry, the SHG from moir\'es requires explicit values for the eighteen independent entries of the $\chi^{(2)}_{ijk}$ tensor [Eqn.~\eqref{eq:chipar}]; see Ref.~\cite{SM}.

Under normal incidence ($\alpha=0^{\circ}$ on the inset of Fig.~\ref{fig:Fig2}(a)), Eqns.~\eqref{eq:aa} and \eqref{eq:ab} have an identical, threefold symmetric dependence on $\phi$. Still considering $\alpha=0^{\circ}$, the SHG expression for the $SP$ hBN bilayer in Eqn.~\eqref{eq:sp} has two-fold symmetry and it peaks at $\phi=90^{\circ}$; the three orientations of $SP-$stacked hBN bilayers related by $120^{\circ}$ and $240^{\circ}$ rotations illustrated in Fig.~\ref{fig:Fig1}(c) help restore a threefold symmetry. Twelve entries of $\chi^{(2)}_{ijk}$ ($\chi^{(2)}_{zzz}$, $\chi^{(2)}_{xxz}$, $\chi^{(2)}_{zxx}$,
 $\chi^{(2)}_{xzz}$, $\chi^{(2)}_{zxz}$,
 $\chi^{(2)}_{yyz}$, $\chi^{(2)}_{zyy}$,
 $\chi^{(2)}_{yzz}$, $\chi^{(2)}_{zyz}$,
 $\chi^{(2)}_{xyz}$, $\chi^{(2)}_{yxz}$, and $\chi^{(2)}_{zxy}$) do not enter the expression for $\chi_{\|}^{(2)}(\theta,\phi,\alpha;-2\omega,\omega,\omega)$ when $\alpha=0^{\circ}$. Figure~\ref{fig:Fig2} illustrates the strong dependence on $\omega$ of the maximum SHG intensity when $\alpha=0^{\circ}$. Similar to Ref.~\cite{Yao2021}, the SHG intensity starts at zero for a centrosymmetric AA' bilayer ($\theta=0^{\circ}$) and it peaks at $60^{\circ}$, regardless of the relative displacement $\boldsymbol{\tau}$; the trend is emphasized as an inset in Fig.~\ref{fig:Fig2}(d).

\begin{figure}[tb]
\includegraphics[width=\columnwidth]{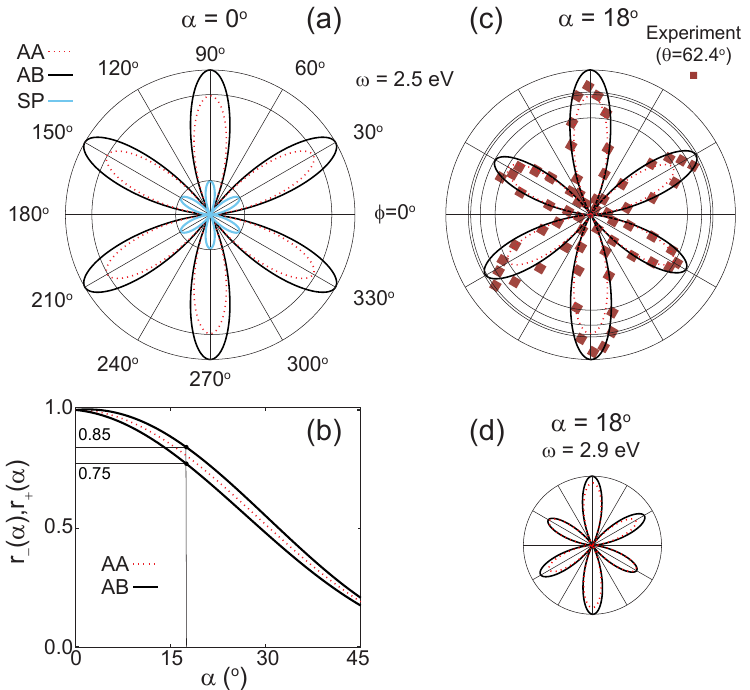}
\caption{(a) Polar dependence of the parallel SHG intensity for $\theta=60^{\circ}$ and $\alpha=0^{\circ}$ under three in-plane displacements $\boldsymbol{\tau}$ leading to $AA$, $AB$, and $SP$ configurations. (b) Inducing a twofold symmetry on the SHG intensity by tilting the probe by $\alpha$; $r_+$ ($r_-$) is the ratio among the largest and second-largest (smallest) lobuli. ($SP$ areas are small domain walls in moir\'es and were not considered for that reason.) (c) The experimental data for $\theta=62.4^{\circ}$ \cite{Yao2021} can be reasonably fitted using $\chi^{(2)}_{||_{AB}}$ and a $\alpha=18^{\circ}$ tilt ($\omega=2.5$ eV). (d) Decrease of the SHG intensity for $\omega=2.9$ eV.  { The radial scale is the same in all plots for direct comparison.}\label{fig:Fig3}}
\end{figure}

Figure~\ref{fig:Fig3}(a) illustrates the polar ($\phi$) sixfold symmetry of the SHG intensity for $AA$, $AB$ ($BA$) and $SP$ ($\theta=60^{\circ}$) hBN bilayers at normal incidence ($\alpha=0^{\circ}$ and $\omega = 2.5$ eV). The experimental polar intensity plot at $\theta=62.4^{\circ}$ in Ref.~\cite{Yao2021} is the closest available data to contrast calculations against, and it lacks sixfold symmetry. Given that the twist angle is so close to $60^{\circ}$, a large-area moir\'e is expected there. (For comparison, a commensurate moir\'e at $\theta=58^{\circ}$ contains 12,944 atoms and 3236 u.c.s \cite{Alejandro}; contrast this to the 21 u.c.s employed in the largest SHG calculations here.)

Under the assumption that a sixfold symmetry is expected in large-area moir\'es at normal incidence, and assuming a coherent and area weighted sum of $\chi^{(2)}_{||_{AA}}$, $\chi^{(2)}_{||_{AB}}=\chi^{(2)}_{||_{BA}}$, and $\chi^{(2)}_{||_{SP}}$ [Fig.~\ref{fig:Fig1}(c)], the experimental data could be explained by the lowering of symmetry of $\chi^{(2)}_{||_{AB}}$ by the presence of the second and third lines in Eqn.~\eqref{eq:ab}, and the presence of the last term in Eqn.~\eqref{eq:sp}, when $\alpha\ne 0^{\circ}$: a tilted incidence changes the maximum intensity of the lobuli of the $\theta=60^{\circ}$ $AB$ ($BA$) bilayers, creating three pairs of lobuli with dissimilar intensities. The ratios $r_+(\alpha)$ and $r_-(\alpha)$ in Fig.~\ref{fig:Fig3}(b) indicate the relative compression of lobuli with respect to those having the largest intensity. When compared to experiment [squares in Fig.~\ref{fig:Fig3}(c)] \cite{Yao2021} , a probe tilted by $\alpha=18^{\circ}$ for $|\chi^{(2)}_{||_{AB}}|^2$ fits the data well. Fig.~\ref{fig:Fig3}(d) shows the strong dependency of the SHG intensity with $\omega$. Indeed, a sizeable decrease of the--twofold symmetric--SHG intensity for $\theta=60^{\circ}$, $\alpha=18^{\circ}$ is documented when the probe frequency $\omega$ is 2.9 eV.

\begin{figure}[tb]
\includegraphics[width=\columnwidth]{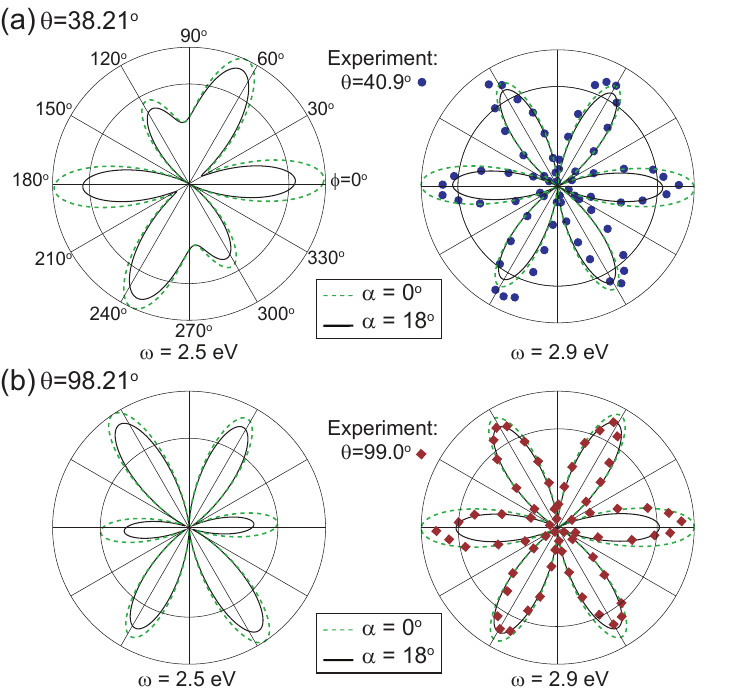}
\caption{Low-symmetry SHG intensity of moir\'es at (a) $\theta=38.21^{\circ}$ and (b) $\theta=98.21^{\circ}$ for $\omega = 2.5$ eV and 2.9 eV, under normal and tilted incidence $\alpha$. Experimental datapoints are from Ref.~\cite{Yao2021}. { The radial scale is the same in all plots for direct comparison.}\label{fig:Fig4}}
\end{figure}

\subsection{SHG from moir\'es}

\begin{figure}[tb]
\includegraphics[width=\columnwidth]{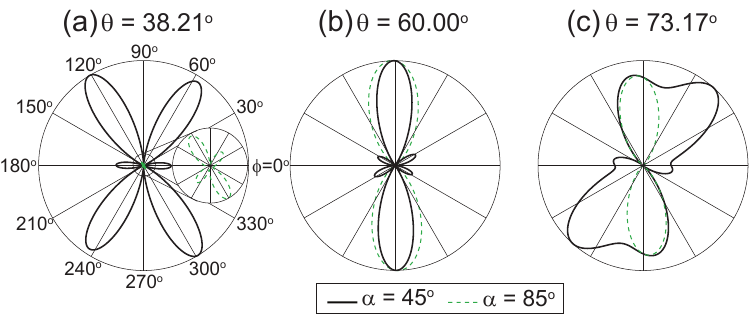}
\caption{SHG intensity for hBN twisted bilayers at (a) $\theta = 38.21^{\circ}$, (b) $\theta = 60.00^{\circ}$ ($AB$), and (c) $\theta = 73.17^{\circ}$ at oblique incidence ($\alpha=45^{\circ}$ and $85^{\circ}$). { The radial scale is the same in all plots for direct comparison.}\label{fig:Fig5}}
\end{figure}

Due to the multiple nonzero entries of the second order susceptibility tensor \cite{SM} because of the low symmetry [Eqn.~\eqref{eq:chipar}], Fig.~\ref{fig:Fig4}(a) shows a SHG intensity of twisted hBN bilayers lacking sixfold symmetry, even when $\alpha=0^{\circ}$, when $\omega=2.5$ eV. The symmetry appears to increase when $\omega=2.9$ eV. Experimental data at $40.9^{\circ}$ is provided, too. Fig.~\ref{fig:Fig4}(b) is a similar plot for $\theta=98.21^{\circ}$: Note the close agreement to experimental data without the need for additional parameters. In particular, note the agreement even for normal incidence ($\alpha=0^{\circ}$). This work ends by showcasing even more exotic SHG intensity profiles. Adjusting the tilt angle $\alpha$ to 45$^{\circ}$, and even 85$^{\circ}$ permits engineering an axial SHG intensity [Fig.~\ref{fig:Fig5}].

\section{Conclusion}
The SHG from twisted hBN bilayers was studied with {\em ab initio} methods, including supercells that contained up to 84 atoms. Moir\'es belong to point group 1, rendering eighteen entries of the second order susceptibility tensor non-zero, and explaining the lack of sixfold symmetry on the SHG intensity in state-of-the-art experiments. In addition, processes to create an axial SHG intensity were elucidated. This work may enable engineered axial sources of SHG and entangled photon pairs from inert and ultrathin 2D insulators.

\acknowledgments

Conversations with A. Pacheco Sanjuan are acknowledged. Work performed in Arkansas was funded by the US Department of Energy (Award No.~DE-SC0022120). Calculations performed at Arkansas were carried out at the Pinnacle Supercomputer, funded by the NSF under award OAC-2346752. BSM thanks the Max Planck Institute for the Structure and Dynamics of Matter
for their hospitality during a Sabbatical stay.


%

\end{document}